\documentclass[a4paper]{jpconf}
\usepackage{subfigure,graphicx} 
\usepackage{amsfonts,amssymb,amsmath}
\usepackage{flafter}
\usepackage{multirow}
\usepackage{txfonts}
\usepackage[unicode,breaklinks]{hyperref}
\usepackage{siunitx}
\newcommand{\papertitle}{Mining the quantum vacuum: quantum tunnelling and particle creation}
\hypersetup{
    unicode=true,
    a4paper=true,
    plainpages=false,
    pdftitle={\papertitle},
    pdfauthor={Ian G. Moss},
    pdfsubject={\papertitle},
    colorlinks=true,
    linkcolor=blue,
    citecolor=blue,
    filecolor=black,
    urlcolor=blue
}

\usepackage{ulem}

\begin{document} 

\title{\papertitle}

\author{Ian G. Moss and Piotr Z. Stasiak}
\address{School of Mathematics, Statistics and Physics, 
Newcastle University, Newcastle upon Tyne, NE1 7RU, UK}
\ead{ian.moss@ncl.ac.uk, P.Stasiak@newcastle.ac.uk}

\date{\today}

\begin{abstract}
Particle production from the vacuum is a remarkable aspect of particle physics.
Prime examples are the Schwinger process of particle production in
strong electric fields and the Hawking process of particle production from black holes.
These processes can be viewed as quantum tunnelling of particles from the vacuum.
The tunnelling approach, and the closely related instanton or complex path approaches,
are reviewed here with emphasis on paths in the complex coordinate plane.
The method is applied to particle production from a black hole in a magnetic field,
where ultra-high energy charged particles are produced.
\end{abstract}

\section{Introduction}

The quantum vacuum is alive with virtual particles that only emerge into reality in extreme
conditions near black holes or in powerful external fields. This particle creation
can be described using various techniques, but the one we focus on here is quantum 
tunnelling from the vacuum. Each methodology has its various strengths, but there are situations 
where the tunnelling approach is especially useful. One particular application where this is the 
case is the production of particles from a magnetic black hole. 

\begin{figure}[htb]
\begin{center}
\scalebox{0.2}{\includegraphics{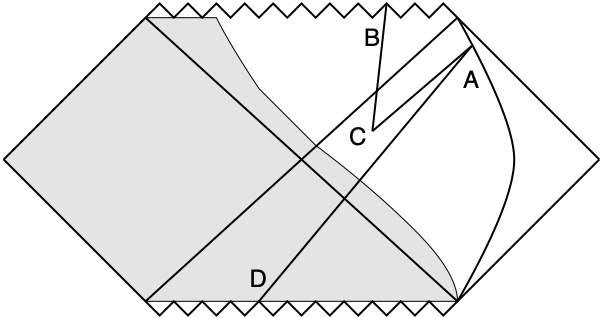}}
\end{center}
\caption{Particle production on a black hole spacetime \cite{PhysRevD.13.2188}. The amplitude for particle
production at the point $C$ can be related to paths $BAC$ and $AD$ by analytic continuation.}
\label{HH}
\end{figure}

The tunnelling approach is influenced by an early description of particle production from black holes 
that appeared in the work of Hartle and Hawking \cite{PhysRevD.13.2188}. They suggested that the amplitude 
for particle production could be related to a particle path from the future singularity to the black hole exterior, 
as in Fig. \ref{HH}. There is no such classical path, but in the analysis, they used analytic continuation of the 
time coordinate to show that the probability $P$ of particle production and absorption for a 
Schwarzschild black hole where related by 
\begin{equation}
P(\hbox{particle emmission})=e^{-\beta E}P(\hbox{particle absorption})\label{rates}
\end{equation}
where $\beta$ is the inverse Hawking temperature. This relation is enough to guarantee
that the black hole can be in equilibrium with a heat bath at the Hawking temperature.

Hartle and Hawking also extended their relation to charged and rotating black holes.
In the period since their pioneering work, analytic continuation has been used to deliver more 
detailed information about the particle production rate beyond the simple relation Eq. (\ref{rates}),
for example with charged black holes \cite{Preparata:1998rz,Kim:2004us,Ruffini1999}. The approach is often employed 
when a quantum field theory approach is problematic, for example for back reaction problems \cite{Parikh:1999mf} 
and for problems with extremal horizons \cite{Angheben:2005rm}. 
The combination of quantum tunnelling and particle pair creation actually preceded the theory of black hole
pair production, first introduced in the context of alternating electric fields \cite{PhysRevD.2.1191}, 
and later developed into a fully consistent theory of the 
Schwinger process \cite{Affleck:1981bma,Kim:2000un,Kim:2003qp,Dunne:2005sx}.
We aim to show that these situations have features in common that 
make it reasonable to refer to them all as quantum tunnelling phenomena.

Astrophysical applications of vacuum breakdown of are somewhat restricted. 
A rotating black hole with a magnetic  field of around $10^{13}{\rm G}$ could in principle 
induce electric field strengths $1.3\times 10^{18}{\rm V}{\rm m}^{-1}$ needed for electron pair creation. 
Holes like this may arise from the collapse of a magnetar to form a black hole, for example \cite{Nathanail:2017wly}. 
However, such systems would be scenes of complex astrophysical phenomena, and secondary pair production
processes from high energy synchrotron photons $\gamma\to e^+e^-$ would likely be prevalent.
Nevertheless, the vacuum production process would generate currents near the horizon and
it may be important to include these in fluid simulations.

Simple estimates of the vacuum breakdown near a black hole can easily be found by taking the
pair creation rates in flat space using the local electric field value in some suitable chosen
reference frame \cite{Preparata:1998rz}. Here we shall improve on this simple approach and 
include the effects of curvature on the particle production. The wave equations for a charged particles 
around a magnetic rotating hole are not separable, but the quantum tunnelling approach
proves invaluable. It turns out that the flat space effect overestimates 
the particle particle production. We shall also be able to determine the 
dynamical parameters of the electrons that are produced and examine their trajectories in some detail. 

The first sections of this paper aims to clarify some of the aspects of particle production using instantons.
In particular, we explore the difference between an instanton that describes vacuum breakdown and an instanton that
describes Hawking radiation from an event horizon. We shall also make extensive use of Hamiltonian methods
and contours in the complex coordinate plane, whose importance for particle production where
extensively studied by Srinivasan and Padmanabhan \cite{Srinivasan:1998ty}

This paper uses a small modification of SI units in which the distance unit is chosen so that
the velocity of light $c=1$.

\section{The instanton approach to quantum tunnelling}

We start with a review of the instanton approach to quantum tunnelling
through a potential barrier, in order to bring out some of the features that will
be important later on. We will introduce Hamilton's principle function and
see how this replaces the usual action, and we will  empahsise the roles
of branch cuts in the complex coordinate plane.

\begin{figure}[htb]
\begin{center}
\scalebox{0.2}{\includegraphics{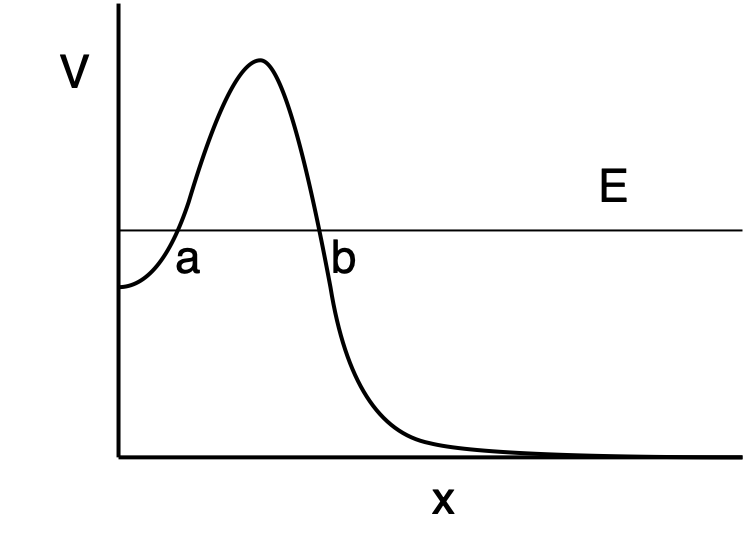}}
\end{center}
\caption{A simple scenario for quantum tunnelling. 
The decay rate is dominated by the WKB approximation with energy $E$
given by an harmonic oscillator state to the left of the barrier.
The exponential factor can be expressed as a contour integral
around the contour $C$.}
\label{TunIns}
\end{figure}

In the simplest situation, a particle tunnels 
from a localised initial state. The particle is prepared at time $t=0$ 
`inside' the barrier, i.e. to the left of the maximum of the the potential $V$ shown in 
figure \ref{TunIns}. The probability of finding the
particle inside the barrier decays exponentially with a rate $\Gamma$, which we
identify as the vacuum decay rate. A simple analysis of the decay rate using the WKB approximation 
to the Schr\"odinger equation gives
\begin{equation}
\Gamma\approx \frac{\omega}{2\pi} \exp\left\{-\frac2\hbar\int_a^b\left\{2m(V-E)\right\}^{1/2}dx\right\}
\label{WKB}
\end{equation}
where $\omega^2=V''(0)/m$ and $E=(n+\frac12)\hbar\omega$ for some integer $n$.

Banks and Bender \cite{PhysRevD.8.3366} demonstrated (in a more general context) 
that the exponent in the decay rate
could be obtained from a classical trajectory $x_b(t_I)$ with imaginary time $t_I=i t$.
The trajectory, or instanton, runs from $x=a$ to $x=b$ in figure \ref{TunIns} and back to $x=a$. 
Consider the classical action
\begin{equation}
S[x]=\int\left\{\frac{m}2\left(\frac{dx}{dt}\right)^2-V\right\}dt.
\end{equation}
Switching to imaginary time,
\begin{equation}
S[x]=i\int\left\{\frac{m}2\left(\frac{dx}{dt_I}\right)^2+V\right\}dt_I.
\end{equation}
Note that, along the instanton trajectory,
\begin{equation}
\frac{m}{2}\left({dx_b\over dt_I}\right)^2-V=-E.\label{cons}
\end{equation}
It is now possible to relate the exponent in the tunnelling rate to the instanton solution. 
First, we introduce Hamilton's principle function $W$,
\begin{equation}
W[x_b]=S[x_b]+E\int_{\cal C} dt,\label{HPF}
\end{equation}
where the contour ${\cal C}$ goes around the path in imaginary time. From Eq. (\ref{cons}),
this can be simplified to
\begin{equation}
W[x_b]=i\int_C\left(V-E\right)dt_I= 2i\int_a^b\left\{2m(V-E)\right\}^{1/2}dx,\label{expo}
\end{equation}
Comparing with the WKB result (\ref{WKB}) gives an important relation
between the tunnelling rate and the principle function,
\begin{equation}
\Gamma\approx \frac{\omega}{2\pi}\exp\left\{-\frac{W_I[x_b]}{\hbar}\right\},
\label{tunnel}
\end{equation}
where $W_I={\rm Im}\,W$. We could stop at this point, but suppose that $V(x)$ is an analytic function, 
then the expression for $W[x_b]$can also be expressed as a contour integral in the complex $x$ plane,
\begin{equation}
W[x_b]=\int_C\left\{2m(E-V)\right\}^{1/2}dx.\label{cexpo}
\end{equation}
In this form, we can distort the contour of integration as long as it goes exactly once around the
branch cut in the integrand. We shall show later that branch cuts and singularities in the complex coordinate
plane play an important role in distinguishing different types of quantum process.

It is useful at this point to compare the result to the theory of vacuum decay \cite{Coleman:1977py}.
Suppose we take Eq. (\ref{tunnel}) and expand in powers of $E/V$. We find,
\begin{equation}
\Gamma\approx A\left(\frac{S_I[x_b]}{2\pi}\right)^{1/2}\exp\left\{-\frac{S_I[x_b]}{\hbar}\right\},
\end{equation}
where $S_I={\rm Im}\,S$ and the factor $A$ depends on the detailed shape of the potential. If we approach the
same problem as a vacuum decay problem, we obtain the same result with the factor
$A$ determined by an operator determinant. Although the two approaches are similar,
we note there are important differences. The result using the function $W$ does not assume 
$E/V$ is small and gives a simpler expression for the factor in front of the exponential
when we have a finite number of degrees of freedom.

In most of the applications considered below we have some ignorable coordinates.
As an example, suppose in the quantum tunnelling problem there are two extra spatial dimensions $y$ and $z$, but
the potential only depends only on $x$. The wave function factorises, and the WKB 
analysis of the tunnelling rate at fixed values of the momenta $p_y$ and $p_z$ is the same as the
one dimensional case. The formula (\ref{tunnel}) is still valid provided we modify the 
definition of the principle function to remove the ignorable coordinates,
\begin{equation}
W=S+Et-y p_y-z p_z.\label{WMod}
\end{equation}
A similar correction should be applied and the modified principle function 
used whenever there are conserved momenta.

\section{The Schwinger process}

The Schwinger process is the pair creation of charged particles, usually electron positron pairs,
in an electric field. Schwinger's original discussion, used heat-kernel methods, and gave
an early example of a non-perturbative result in quantum field theory. We shall review the tunnelling 
approach to the Schwinger process with the aim of obtaining some general rules for the 
tunnelling instanton.

\begin{center}
\begin{figure}[htb]
\begin{center}
\scalebox{0.4}{\includegraphics{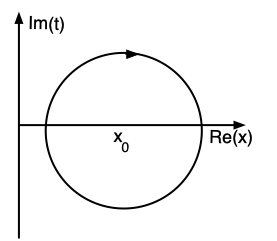}}
\scalebox{0.4}{\includegraphics{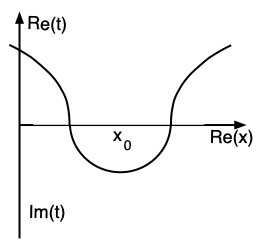}}
\end{center}
\caption{The instanton for the Schwinger process and Vilenkin's `ex nihilo' version.
}
\label{SchIns}
\end{figure}
\end{center} 

Consider a particle with mass $m$ and charge $e$.
The particle world-line $x^\mu(\tau)$ is parameterised by proper time $\tau$.
The action can be expressed in Hamiltonian form with momenta $p_\mu$,
\begin{equation}
S=\int\left(\dot x^\mu p_\mu-H\right)d\tau.
\end{equation}
Given the metric $g_{\mu\nu}$, the vector potential $A_\mu$ and charge $e$,
\begin{equation}
H=\frac{1}{2m}g^{\mu\nu}(p_\mu+eA_\mu)+\frac{m}{2}.\label{Hgen}
\end{equation}
We take flat spacetime with a constant electric field ${\cal E}$ in the $x$ direction, 
associated with a potential $A_t={\cal E}x$.  The resullting Hamiltonian is
\begin{equation}
H=-\frac1{2m}\left(p_t+e{\cal E}x\right)^2+\frac1{2m}p_x^2+\frac1{2m}p_y^2+\frac1{2m}p_z^2+\frac{m}{2}.
\end{equation}
Normalisation of four-velocity $\dot x^\mu$  imposes a constraint $H=0$ on the Hamiltonian.
Furthermore, ignorable coordinates $t$, $y$ and $z$ imply that the 
energy $E=-p_t$ and momenta $p_y$, $p_z$ are conserved. With these restrictions, 
the modified principle function (\ref{WMod}) reduces to
\begin{equation}
W=S+Et-y p_y-z p_z=\int p_x dx\label{wgen}
\end{equation}
For convenience, we introduce a new parameter $x_0$ related to the energy by $E=e{\cal E}x_0$, then
the constraint $H=0$ implies
\begin{equation}
p_x^2=(e{\cal E})^2\left(x-x_0\right)^2-m^2-p_\perp^2,\label{px}
\end{equation}
where $p_\perp$ is the momentum perpendicular to the $x$ direction. 
Note that, for real values of position $x$, we take
the positive square root for $p_x$.

In the tunnelling approach, we evaluate the tunnelling exponent Eq. (\ref{wgen}) for a solution of the equations of motion
that runs along a closed complex contour ${\cal C}$ in the complex $x$ plane. The tunnelling exponent 
using Eq. (\ref{px}) is,
\begin{equation}
W=\int_{\cal C}\left((e{\cal E})^2\left(x-x_0\right)^2-m^2-p_\perp^2\right)^{1/2}dx,\label{Wint}
\end{equation}
The integrand has a branch cut between $x_0\pm\kappa$, where $\kappa=(m^2+p_\perp^2)^{1/2}/|e{\cal E}|$. 
In order to find a suitable integration contour we start from the general solution to
the equations of motion in  real time,
\begin{align}
x-x_0&=\kappa\cosh\frac\tau\kappa\label{xInst}\\
t-t_0&=\kappa\sinh\frac\tau\kappa,
\end{align}
Consider the complex contour
\begin{equation}
\tau=-i\kappa\phi+\epsilon,
\end{equation}
where the real parameter $\phi$ lies on a circle
and an $i\epsilon$ prescription has been used to avoid the branch cut in Eq. (\ref{Wint}).
In the plane with axes ${\rm Re}(x)$ and ${\rm Im}(t)$, the contour is a circle, as shown
in Fig. \ref{SchIns}. The direction has been chosen so that the principle value
of the square root will result in a positive imaginary part for the integral. 
If we use the negative root in Eq. (\ref{Wint}) then we take a counter-clockwise contour.
An interesting interpretation of the instanton has been suggested by Vilenkin \cite{Vilenkin:1982de}.
Combining the bottom half of the instanton  to the real time evolution of the particle worldlines
for $t>0$ produces the picture on the right. From the point of view of an observer in real time,
the electron positron pair suddenly appears as if we have `creation from nothing'.
Strange behaviour should be expected when we try to interpret a quantum phenomenon in 
purely classical terms.

\begin{center}
\begin{figure}[htb]
\begin{center}
\scalebox{0.4}{\includegraphics{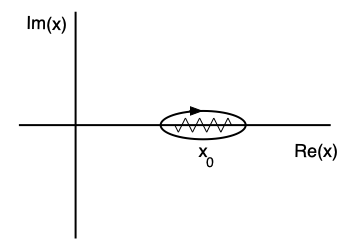}}
\scalebox{0.4}{\includegraphics{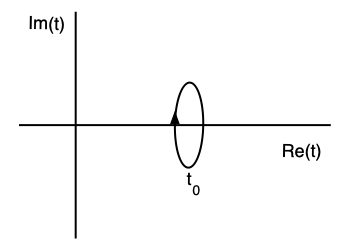}}
\end{center}
\caption{Alternative views of the instanton for the Schwinger process. The
complex space plane (left) and the complex time plane (right).
}
\label{SchInsB}
\end{figure}
\end{center} 

The first diagram in Fig. \ref{SchInsB} shows how the instanton contour goes around the branch 
cut in the complex $x$ plane. Note that any contour which circles the branch cut 
clockwise exactly once gives the same value of the tunnelling rate, so that the only
ambiguity in the result lies in the winding number of the contour.
The second diagram in Fig. \ref{SchInsB} shows the contour in the complex
$t$ plane. In this picture the tunnelling contour can be split into a
particle line and an antiparticle line. Each line contributes half of the closed
instanton path, and an instanton with winding number $-n$ would represent the
production of $n$ particle-antiparticle pairs.

Integrating (\ref{Wint}) along the contour around the branch cut gives the tunnelling exponent,
\begin{equation}
\frac{W_I}{\hbar}=\frac{\pi m^2}{|e{\cal E}|\hbar}+\frac{\pi p_\perp^2}{|e{\cal E}|\hbar}.\label{SW}
\end{equation}
The prefactor for the tunnelling rate in the barrier penetration case was $(V''/2\pi m)^{1/2}$, and we will
divide this by the Compton wavelength $\hbar/mc$ to get the correct dimensions.
Putting in a phase space factor in addition gives an estimate for the particle production
$d\Gamma$ with transverse momentum $p_\perp$,
\begin{equation}
d\Gamma=\frac{|e{\cal E}|}{2\pi\hbar}\left(\frac{dp_\perp}{2\pi\hbar}\right)^2e^{-W_I/\hbar}
\end{equation}
After integrating the particle production rate $d\Gamma$ over the transverse momenta, we obtain
the correct formula for the particle production rate $\Gamma$ per unit volume \cite{Kim:2000un} ,
\begin{equation}
\Gamma={1\over \pi}\left({e{\cal E}\over 2\pi\hbar}\right)^2e^{-\pi m^2/\hbar  |e{\cal E}|}\label{srate}
\end{equation}
An exponent $\pi m^2/\hbar  |e{\cal E}|\approx 1$ for electrons corresponds to an electric field strength 
$4.157\times10^{18}\,{\rm V}{\rm m}^{-1}$.  Pair production is heavily suppressed for smaller
field strengths. On the other hand, ordinary perturbation theory can be used to describe pair production for 
larger field strengths. The result is only useful over a limited range of field strengths.

In conclusion, the Schwinger process is represented by a closed contour
around a branch cut in the complex coordinate plane. Two halves of the contour with single
winding number in the complex time plane represent production of a particle and an antiparticle.

\subsection{The thermal Schwinger process}

The production of particles in an electric field at finite temperature  gives another application
of the tunnelling approach \cite{Medina:2015qzc,Gould:2018ovk,Gould:2018efv}.  
Thermal tunnelling rates in quantum mechanics are related to the imaginary part of the
free energy \cite{Affleck:1980ac}. In the path integral approach, we find the free energy
by imposing a periodicity $\beta=\hbar/k_BT$ on the action in imaginary time.
We do the same for calculating the particle creation rate. 
As before, the the main focus here will be on the choice of contour for the instanton
approximation.

\begin{center}
\begin{figure}[htb]
\begin{center}
\scalebox{0.4}{\includegraphics{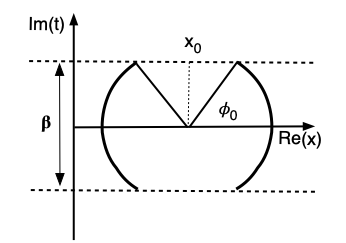}}
\scalebox{0.4}{\includegraphics{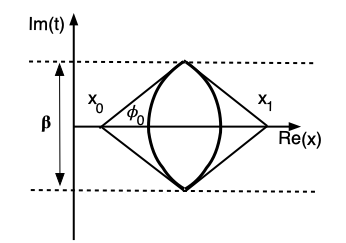}}
\end{center}
\caption{The instanton for the thermal Schwinger process. Pasting the circular instanton
on to the periodic manifold gives a non-differentiable path (left). Moving the pieces together
gives the differentiable path (right).
}
\label{SchTemp}
\end{figure}
\end{center} 

In the Schwinger process, the periodicity in imaginary time cuts off
the top and bottom of the circular instanton as shown in figure \ref{SchTemp}.  The contour would be
continuous on the periodic manifold,  but not differentiable. In order to obtain a differentiable 
contour we move the left and right segment together as on the right side of figure \ref{SchTemp}. 
This adjustment is essential for obtaining the correct value of the instanton action.

The right segment is centred at $x_0$ and the left segment at $x_1$. The corresponding
integrals are denoted by $W_R$ and $W_L$, and evaluated using
the angle $\phi=i\tau/\kappa$ as independent variable in Eqs. (\ref{Wint}) and (\ref{xInst}). 
The contributions are
\begin{align}
W_R&=im\kappa \int_{-\phi_0}^{\phi_0}\sin^2\phi\,d\phi
-imx_0\int_{-\phi_0}^{\phi_0}\cos\phi\,d\phi\\
W_L&=im\kappa \int_{\pi-\phi_0}^{\pi+\phi_0}\sin^2\phi\,d\phi
-imx_1\int_{\pi-\phi_0}^{\pi+\phi_0}\cos\phi\,d\phi,
\end{align}
where $\sin\phi_0=\beta/2\kappa$.
The final result is independent of $x_0$ and $x_1$ because of the
identity $x_0-x_1=2\kappa\cos\phi_0$. The total functon $W=W_R+W_L$,
\begin{equation}
W=2i\kappa m\left\{\phi_0+\frac12\sin2\phi_0\right\}
\end{equation}
The tunnelling exponent \cite{Medina:2015qzc},
\begin{equation}
\frac{W_I}{\hbar}=\frac{2\kappa m}{\hbar}\left\{\arcsin\left(\frac{\beta}{2\kappa}\right)+
\frac{\beta}{2\kappa}\left[1-\frac{\beta^2}{4\kappa^2}\right]^{1/2}\right\},
\end{equation}
where $\kappa=(m^2+p_\perp^2)^{1/2}/|e{\cal E}|$.
This reproduces the Schwinger result (\ref{SW}) in the zero temperature limit. In the high
temperature limit, $W_I/\hbar\to 2\beta m$,  which represents the probability of finding a 
particle-antiparticle pair at high temperature. In future, whenever we see 
$W_I/\hbar\to \beta E$ we will interpret this as a signal of thermal particle production at temperature
$k_BT=\hbar/\beta$.

\section{The Fulling-Davies-Unruh effect}

The next example is the detection of thermal particles by an accelerating detector.
We take the detector to be at rest in a two-dimensional accelerating frame with acceleration $g$.
We shall review the tunnelling description to see what features of tunnelling instantons
are typical of thermal particle production.

The accelerating frame is associated with a set of Rindler coordinates $(x,t)$,
and metric  
\begin{equation}
ds^2=-g^2 x^2 dt^2+dx^2.
\end{equation}
The Hamiltonian (\ref{Hgen}) for particle motion $x(\tau)$ and $t(\tau)$,  is
\begin{equation}
H=-\frac1{2m}\frac{p_t^2}{g^2x^2}+\frac1{2m}p_x^2+\frac{m}{2}.
\end{equation}
The energy $E=-p_t$ is conserved and the Hamiltonian is constrained to $H=0$. 
As before, the tunnelling is related to the principle
function integrated around a closed contour,
\begin{equation}
W=\int_{\cal C} p_x dx,
\end{equation}
Using the Hamiltonian constraint,
\begin{equation}
W=\int_{\cal C} \left(\frac{E^2}{g^2}-m^2x^2\right)^{1/2} \frac{dx}{x}
\end{equation}
To investigate the integration contour, we take the general solution to the equations
of motion,
\begin{align}
x&=\left(\frac{E^2}{m^2g^2}-\tau^2\right)^{1/2},\label{FDcontour}\\
t-t_0&=\frac1{g}\log\left(\frac{E+mg\tau}{E-mg\tau}\right)^{1/2}.
\end{align}
Consider proper time contour
\begin{equation}
\tau=-\frac{E}{mg}+\epsilon e^{2i\phi},
\end{equation}
where $\phi$ lies on the circle. This gives a circular contour in the complex 
$x$ plane around the horizon $x=0$. Any closed contour which goes around
the horizon singularity once will give the same value for the tunnelling exponent.
In the complex $t$ plane, the contour goes between $t_0\pm 2\pi i/g$. The metric
is regular and {\it the contour is closed if we impose periodicity of the metric in 
imaginary time}.

\begin{center}
\begin{figure}[htb]
\begin{center}
\scalebox{0.4}{\includegraphics{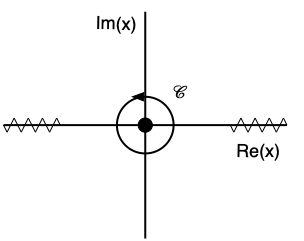}}
\scalebox{0.4}{\includegraphics{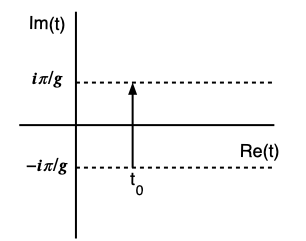}}
\end{center}
\caption{Two views of the instanton for the Unruh process. The complex space plane (left) and
the complex time plane (right). 
}
\label{KerrIns}
\end{figure}
\end{center}

Integrating around the singularity using the residue theorem gives exponent
\begin{equation}
\frac{W_I}{\hbar}=\frac{2\pi E}{\hbar g}.
\end{equation}
This has the thermal interpretation $W_I/\hbar=E/k_BT$ as in relation (\ref{rates}), 
where the Unruh temperature
\begin{equation}
T=\frac{\hbar g}{2\pi k_B}.
\end{equation}

We should also examine what happens if we use a different coordinate system, specifically
putting the metric in Boyer-Lindquist form with $r=gx^2/2$,
\begin{equation}
ds^2=-fdt^2+f^{-1}dr^2, \hbox{ where }f=g^2 r^2.
\end{equation}
The same contour (\ref{FDcontour}) which wound once around the horizon in the complex $x$ plane
now winds {\it twice} around the horizon in the complex $r$ plane, though the periodicity in the
complex $t$ plane and the particle production rate remain the same.

In conclusion, horizon radiation is represented by a closed contour 
around a singularity in the complex coordinate plane. The contour is closed in the 
complex time plane only when we impose impose periodicity in imaginary time.
The contour has winding number two in the complex $r$ plane when we use the 
Boyer-Lindquist coordinates.

\section{Charged black holes}

Radiation from charged black holes can include contributions from thermal radiation
with the Hawking temperature $T_h$ and breakdown of the vacuum
due to the electric field outside the black hole.
All forms of radiation are included in the simple expression
for the particle flux obtained from a mode decomposition of the Dirac or the wave equation
\cite{Gibbons:1975kk},
\begin{equation}
F=\sum_{l,m}\int_0^\infty \frac{d\omega}{2\pi}\left(1-|A_{lm}|^2\right)\frac{1}{e^{\beta\omega_h}-1},
\end{equation}
where the inverse temperature $\beta=\hbar/T_H$. The frequency 
$\omega_h=\omega-e\Phi_h$, where $\Phi_h$ is the electrostatic potential at the horizon.
The amplitude $A_{lm}$ represents reflection of the particle modes with angular wave numbers $l$ and $m$
back into the black hole. This amplitude can only be obtained numerically, or using approximate methods for various regimes.

The quantum tunnelling approach to particle creation can be used to obtain closed expressions
in the regime $|\beta\omega_h|\gg1$, when it is related to using WKB approximations to
the reflection amplitude. In this limit,
the flux integral can be decomposed into two parts:

\noindent{\it The super-radiant regime} $\omega<e\Phi_h$ where the flux  becomes
\begin{equation}
F_{\rm super}\approx\sum_{l,m}\int_0^{e\Phi_h}\frac{d\omega}{2\pi}\left(|A_{lm}|^2-1\right)\label{SR}
\end{equation}
It is in this regime that electromagnetic breakdown of the vacuum can occur.

\noindent{\it The non-super-radiant regime} $\omega>e\Phi_h$, where
\begin{equation}
F_{\rm thermal}\approx\sum_{l,m}\int_{e\Phi_h}^\infty\frac{d\omega}{2\pi}
\left(1-|A_{lm}|^2\right)e^{-\beta\omega_h},\label{NSR}
\end{equation}
which we can regard as the Maxwell-Boltzmann approximation to the thermal Hawking flux
filtered by a grey-body factor. We shall now show how the quantum tunnelling approach
reproduces these results.

\subsection{The tunnelling approach}

The spacetime is described by the Reissner-Nordstrom metric
\begin{equation}
ds^2=-fdt^2+f^{-1}dr^2+r^2(d\theta^2+\sin^2\theta d\phi^2),
\end{equation}
where
\begin{equation}
f=1-\frac{2M}{r}+\frac{Q^2}{r^2}.
\end{equation}
Geometric mass $M=GM_*$ and Geometric charge
$Q=G^{1/2}Q_*/(4\pi\epsilon_0)^{1/2}$ are related to the physical mass and charge $M_*$ and $Q_*$.
The electrostatic potential $\Phi$ at radius $r$ is
\begin{equation}
\Phi=\frac{Q_*}{4\pi\epsilon_0 r}.
\end{equation}

\begin{center}
\begin{figure}[htb]
\begin{center}
\scalebox{0.4}{\includegraphics{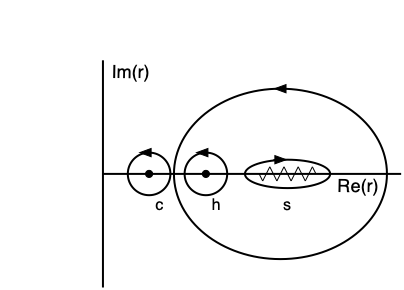}}
\end{center}
\caption{The complex $r$ plane showing different contours.
}
\label{RNIns}
\end{figure}
\end{center} 

Due to the rotational symmetry, it will be sufficient to start from the Hamiltonian (\ref{Hgen}) 
for a particle of charge $e$ in the equatorial plane, with conserved momenta $E=-p_t$ and $L=p_\phi$,
\begin{equation}
H=-\frac{(E-e\Phi)^2}{2mf}+\frac{fp_r^2}{2m}+\frac{1}{2m}\left(m^2+\frac{L^2}{r^2}\right).
\end{equation}
The modified principle function $W=S+Et-L\phi$ is,
\begin{equation}
W=\int_{\cal C}p_rdr=\int_{\cal C} 
\frac{1}{f}\left\{\left(E-e\Phi\right)^2-\left(m^2+\frac{L^2}{r^2}\right)f\right\}^{1/2}dr.
\end{equation}
There are poles at the outer and inner horizons $r_h$ and $r_c$, as well as possible branch cuts. A typical
representation of the complex $r$ plane is shown in figure \ref{RNIns}. The contribution from each
of the contours will be denoted by a subscript, e.g. $W_\infty$ for the large outer contour.

From the large radius limit, we find
\begin{equation}
W_\infty=\frac{4\pi iM}{(E^2-m^2)^{1/2}}
\left\{ \left(E^2-\frac{e\Phi_hr_h}{2M}\right)-\frac{m^2}{2}\right\}
\end{equation}
The horizon integrals are obtained from the residue theorem,
\begin{align}
W_h&=\frac{\pi i}{\kappa_h}\left(E-e\Phi_h\right)\\
W_c&=\frac{\pi i}{\kappa_c}\left(E-e\Phi_c\right)
\end{align}
where the surface gravities $\kappa_h=f'(r_h)/2$ and $\kappa_c=f'(r_c)/2$.
Integrals around the branch cuts can be deduced from the other integrals using Cauchy's theorem.

\subsection{Black hole Schwinger process}

The Schwinger process for electron-positron production is represented by a contour 
which goes around the branch cut. This contribution is independent of the Hawking temperature
and we identify it with the super-radiant flux (\ref{SR}). From Cauchy's theorem, the 
principle function $W_S=W_h+W_c-W_\infty$. The imaginary part,
\begin{equation}
W_I=\frac{4\pi M}{(E^2-m^2)^{1/2}}
\left\{\left(E-(E^2-m^2)^{1/2}\right)\left(E-\frac{e\Phi_hr_h}{2M}\right)-\frac{m^2}{2}\right\}.
\end{equation}
In the large energy limit, the tunnelling exponent at leading order of $m/E$ is
\begin{equation}
\frac{W_I}{\hbar}=\frac{\pi m^2 e\Phi_hr_h}{\hbar E^2}.
\end{equation}
The angular momentum $L$ only appears in the location of the branch cut. If $L\ll Er_h$, then
the branch cut is narrow with centre at the radius where $E-e\Phi=0$. Physically, this represents
the radius $r(E)$ at which the particles of energy $E$ are created.
The electric field at the centre of the branch cut is
\begin{equation}
e{\cal E}=\frac{e\Phi}{r}=\frac{E^2}{e\Phi_hr_h}\label{locE}
\end{equation}
Hence
\begin{equation}
\frac{W_I}{\hbar}=\frac{\pi m^2}{\hbar e\cal E}.\label{wicharged}
\end{equation}
This recovers the Schwinger result, but with the local electric field ${\cal E}$ at the radius where
the particles are created. We conclude that the particle production is sufficiently localised for the 
equivalence principle to hold. Furthermore,
we can use the Schwinger result to infer the pre-factor for the particle production rate per unit volume,
\begin{equation}
\Gamma={1\over \pi}\left({e{\cal E}\over 2\pi\hbar}\right)^2e^{-\pi m^2/e\hbar  {\cal E}}.\label{bhsrate}
\end{equation}
As with the flat spacetime result, this is only valid for large electric fields.

\begin{center}
\begin{figure}[htb]
\begin{center}
\scalebox{0.3}{\includegraphics{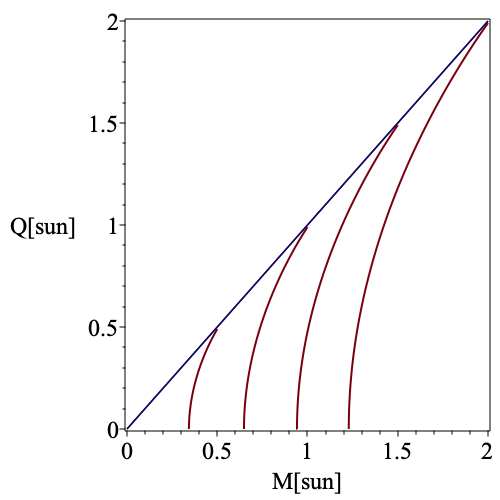}}
\end{center}
\caption{Evolution of the geometric mass and charge parameters due to particle production
 is downwards along the red lines in this plot. Thermal emission is tiny and not included.
 Units are solar Schwartzchild radii
 (2.9 km).
}
\label{QMPlot}
\end{figure}
\end{center} 

The total luminosity of the black hole can be obtained by integrating the particle production
for the region outside of the horizon. Because of the relation between the location of
particle creation $r$ and the energy, this is equivalent to integrating over the energy.
First, we rewrite the particle production rate in terms of radius $r$ using (\ref{locE}) and (\ref{wicharged}),
\begin{equation}
\Gamma={1\over 4\pi\alpha^2}\left({mr_h\over \hbar r}\right)^4e^{-\alpha r^2/r_h^2},\label{bhsrate}
\end{equation}
where
\begin{equation}
\alpha=\frac{\pi m^2r_h}{\hbar e\Phi_h}.
\end{equation}
The evaporation rate is then
\begin{equation}
\frac{dM_*}{dt}=\int_{r_h}^\infty  dr\,4\pi r^2\,\Gamma E=
\frac{\pi}{\alpha^2 r_h}\left( \frac{mr_h}{\hbar} \right)^5\Gamma(-2,\alpha), 
\end{equation}
where $\Gamma(a,x)$ is the incomplete Gamma function.
The charge evaporates at a rate
\begin{equation}
\frac{dQ_*}{dt}=\int_{r_h}^\infty  dr\,4\pi r^2\,\Gamma e
=\frac{\pi e}{\alpha^{3/2} r_h}\left( \frac{mr_h}{\hbar} \right)^4\Gamma(-3/2,\alpha)
\end{equation}
The relative rates of (geometric) charge and mass evaporation has a simple expression,
\begin{equation}
\frac{dQ}{dM}=\frac{Q}{r_h}\frac{\Gamma(-2,\alpha)}{\alpha^{1/2}\Gamma(-3/2,\alpha)}
\end{equation}
It is a known result that the black hole looses charge due to super-radiance
at a far higher rate than it looses mass \cite{Hiscock:1990ex}. However, having an expression in closed form
is a success of the tunnelling approach.

\subsection{Black hole Hawking process}

The Hawking flux has two contributions. For $E>e\phi_h$, there is a contribution
from the contour which circles the horizon and represents particle production
at the horizon.  We may also have contributions from branch cut outside the horizon
which now represents the transmission term $|A_{lm}|^2$ through the potential barrier.
The horizon contribution has winding number two in the coordinate system in use,
as we saw earlier in the context of the Fulling-Davies-Unruh effect. 
The integral $W_H=2W_h$ gives a particle creation rate
\begin{equation}
\Gamma_H\propto e^{-\beta(E-e\Phi_h)/\hbar}
\end{equation}
which agrees with the first term in (\ref{NSR}), at the Hawking temperature $T_H=\hbar\kappa_h/2\pi$.

\section{Particle production on a magnetic rotating black hole background}

In this section we apply the tunnelling method to the production of electron-positron pairs from
the vacuum around a rotating black hole in an external magnetic field. The Hawking radiation
is insignificant for large black holes, and so with astrophysical applications in mind we consider 
only the Schwinger process. However, we take an idealised vacuum situation with no other 
particles present.

\subsection{Geometry}

For a solar-mass black hole, the back-reaction of the magnetic field on the geometry is small
when $B\le 10^{15}\,G$ and the Kerr metric can be used,
\begin{equation}
ds^2=-\frac{\Delta}{\rho^2}\omega^{t\ 2}+\frac{s^2}{\rho^2}\omega^{\phi\ 2}+\frac{\rho^2}{\Delta}dr^2
+\rho^2d\theta^2,
\end{equation}
where
\begin{equation}
\omega^t=dt-as^2d\phi,\qquad \omega^\phi=(a^2+r^2)d\phi-adt,
\end{equation}
The metric functions are $\Delta=r^2+a^2-2Mr$, $s=\sin\theta$ and $\rho^2=r^2+a^2-a^2s^2$.
The geometric mass $M$ is $G/c^2$ times the physical mass. 

We will take a magnetic field with rotational symmetry about the black hole axis and assume
the simplest dipole field that approaches a constant field with strength $B$ in the $z$ direction at large distances.
Furthermore, we will assume that the movement of charged particles leaves the black hole with a net charge that 
neutralises the electromotive force (EMF).  The electromagnetic potential for this zero EMF field has components
\begin{equation}
A_t=-\frac{Mars^2}{\rho^2}B,\qquad A_{\phi}=\frac{A s^2}{2\rho^2}B,
\end{equation}
where $A=(r^2+a^2)^2-\Delta a^2 s^2$.
Although the EMF vanishes, there is an electric field in the non-rotating (zero angular momentum)
frame defined in Ref. \cite{Bardeen:1972fi}. We shall see that this electric field is associated with
the particle production.

\subsection*{Dynamics}

Some basic dynamical notions will be needed for the particle production calculation.
The four-momentum $p_\mu$ for a particle with mass $m$, charge $e$ and four velocity $u^\mu$ is
\begin{equation}
p_\mu=mg_{\mu\nu}\left(u^\nu+e A^\nu\right)
\end{equation}
Along the Killing directions, we set
\begin{equation}
E=-p_t,\qquad L=p_\phi,
\end{equation}
The momenta are related by the constraint
\begin{equation}
g^{\mu\nu}p_\mu p_\nu=-m^2,
\end{equation}
After inserting the metric components,
\begin{equation}
\frac{\Delta}{\rho^2}p_r^2+\frac{1}{\rho^2}p_\theta^2+V=0,\label{Econs}
\end{equation}
where the effective potential $V$ is given by
\begin{equation}
V=-\frac{A}{\Delta \rho^2}\left( E-\Omega L\right)^2
+\frac{\rho^2}{As^2}\left(L-\frac{eBAs^2}{2\rho^2}\right)^2+m^2.
\label{Vpot}
\end{equation}
The local rotation rate $\Omega=2Mra/A$.

\subsection{Tunnelling exponents}

The tunnelling exponent is given by ${\rm Im} W/\hbar$, 
where the modified principle function  $W=S+Et-L\phi$.
Inserting the action leaves
\begin{equation}
W=\int_{\cal C} \left(p_r dr+p_\theta d\theta\right).
\end{equation}
Unlike in the previous examples, there are two remaining coordinates
$r$ and $\theta$, but both are implicitly functions of the proper time $\tau$.
The complex contour $C$ for the Schwinger process surrounds a branch
cut and gives an imaginary value to $W$. This happens in a region
where classical trajectories are forbidden because $V$ is negative,
and the momenta are therefore complex. Tunnelling occurs
inside a potential barrier that ends at points PQ as shown in Fig \ref{KerrIns}.

\begin{center}
\begin{figure}[htb]
\begin{center}
\scalebox{0.4}{\includegraphics{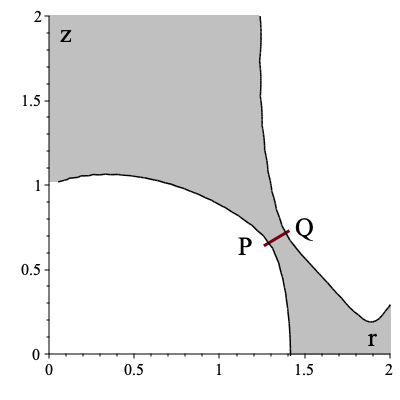}}
\end{center}
\caption{A region of the $(r,\theta)$ plane showing regions of positive potential (grey) and
negative potential (white). The line PQ shows the trajectory of an instanton, with the particle-antiparticle
pair produced at P and Q.
}
\label{KerrIns}
\end{figure}
\end{center} 

A crucial observation is that the tunnelling only occurs with any significant rate for very small values of $W_I$ 
compared to the astrophysical scales set by the mass of the black hole. This requires both brackets
in the potential (\ref{Vpot}) to be very small, and restricts the values
of the energy and angular momentum. The centre of the barrier $r=r_c$, $\theta=\theta_c$ is located where
both brackets vanish,
\begin{equation}
E=\Omega_c L,\qquad L=\frac{eBs_c^2A_c}{2\rho^2_c}\label{EandL}
\end{equation}
These relate both the energy and angular momentum to $r_c$ and $s_c=\sin\theta_c$.
Because the barrier is extremely narrow, we can think of pair creation for particles with energy $E$
happening along the circle at $r_c(E)$ and $\theta_c(E)$.

In the $x^i=(r,\theta)$ sector, the Hamiltonian that generates the field equations is
\begin{equation}
H=\frac{1}{2m}g^{ij}p_ip_j+\frac{1}{2m}V
\end{equation}
In the region of the barrier, we introduce small quantities $\delta q^1=r-r_c$ and $\delta q^2=\theta-\theta_c$,
and we use a quadratic approximation to the Hamiltonian,
\begin{equation}
H=\frac{1}{2m}g^{ij}p_i p_j+\frac{1}{4m}V_{,ij}\delta q^i\delta q^j+\frac{m}{2},
\end{equation}
where the Hessian of the potential is evaluated at the centre of the barrier $(r_c,\theta_c)$.
We diagonalise the Hamiltonian by solving the eigenvalue problem for basis vectors $e_n^i$,
\begin{equation}
V_{,ij} e_n^j=2m^2 \lambda_n g_{ij} e_n^j.\label{evp}
\end{equation}
Introduce normal mode coordinates $x^n$, where
\begin{equation}
\delta q^i=x^n\,e_n^i\label{normalmode}
\end{equation}
In terms of the normal modes,
\begin{equation}
H=\frac{1}{2m}\delta^{mn} p_mp_n+\frac12m\Lambda_{mn}x^mx^n+\frac{m}{2},
\end{equation}
where $\Lambda_{mn}={\rm diag}(\lambda_1,\lambda_2)$.
For a compact instanton, we must use the mode $x$ which has a negative 
eigenvalue $\lambda=-\omega^2$. This is the mode that corresponds to the line PQ in figure \ref{KerrIns}.
For this mode,
\begin{equation}
H=\frac{1}{2m}p_x^2-\frac12m\omega^2 x^2+\frac{m}{2}=0
\end{equation}
The principle function is
\begin{equation}
W=\int_{\cal C} p_xdx=im\int_{\cal C}(1-\omega^2 x^2)^{1/2}dx,
\end{equation}
where the contour winds once around the branch cut for a single pair creation event. Hence
\begin{equation}
W=\frac{im\pi}{\omega}.
\end{equation}
A better feel for the result can be obtained by scaling out the dimensionfull quantities from $\omega$,
\begin{equation}
\omega=\frac{eB}{m}\hat\omega,
\end{equation}
where $\hat\omega\equiv\hat\omega(r_c/M,a/M,\theta_c)$ is dimensionless, and obtained by
solving the eigenvalue problem (\ref{evp}) with $\lambda=-\omega^2$.
The particle production rate at $(r_c,\theta_c)$ is $\propto \exp(-W_I/\hbar)$, where
\begin{equation}
\frac{W_I}{\hbar}=\frac{\pi m}{\hbar \omega}=\frac{\pi m^2}{\hbar \hat\omega eB}={\pi B_0\over \hat\omega B},\label{bhB}
\end{equation}
and $B_0=m^2/e\hbar=4.4\times 10^{13}\,G$ for electrons.
In general, the factor $\hat\omega$ has to be obtained numerically, but in the special case of equatorial
particle production the value has a closed form,
\begin{equation}
\hat\omega=\frac{1}{r^2}\left\{a^2(r+M)^2-\Delta r^2\right\}^{1/2}.
\end{equation}

\subsection{Particle fluxes}

\begin{figure}[htb]
\begin{center}
\scalebox{0.4}{\includegraphics{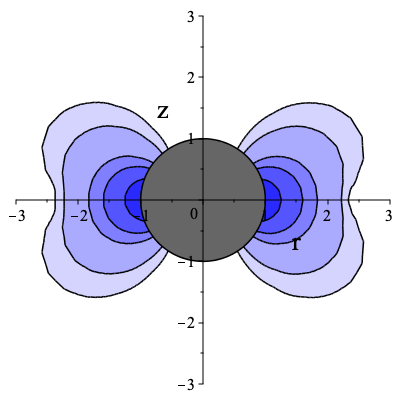}}
\scalebox{0.4}{\includegraphics{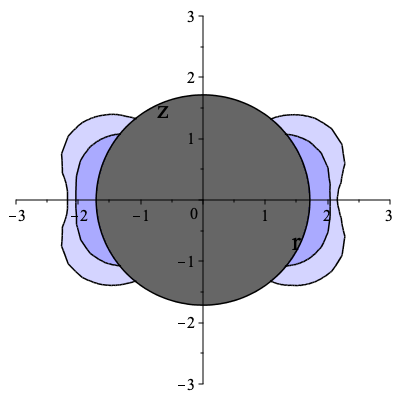}}
\end{center}
\caption{A region of the $r$ and $\theta$ plane showing contours of constant particle production rate.
The colours represent the factor $\pi/\hat\omega$. The extremal case
$a=M$ on the left and $a=0.7M$ on the right.
}
\label{Flux}
\end{figure}

The factor $\pi/\hat\omega$ that determines the particle production rate has been plotted in figures \ref{Flux}, 
where we see the the relative amount of particle production for different values of $r$ and $\theta$.
Particle production is concentrated close to the horizon. 
Initially, $p_r=p_\theta=0$, and the the particles move in circular orbits. As more particles are produced, 
a current loop will build up which produces
a field counteracting the original field. Gradually, the instability  in the normal mode `$x$' direction drives particles away 
from their circular orbits, and into a chaotic ones \cite{Takahashi:2008zh}.

\begin{figure}[htb]
\begin{center}
\scalebox{0.4}{\includegraphics{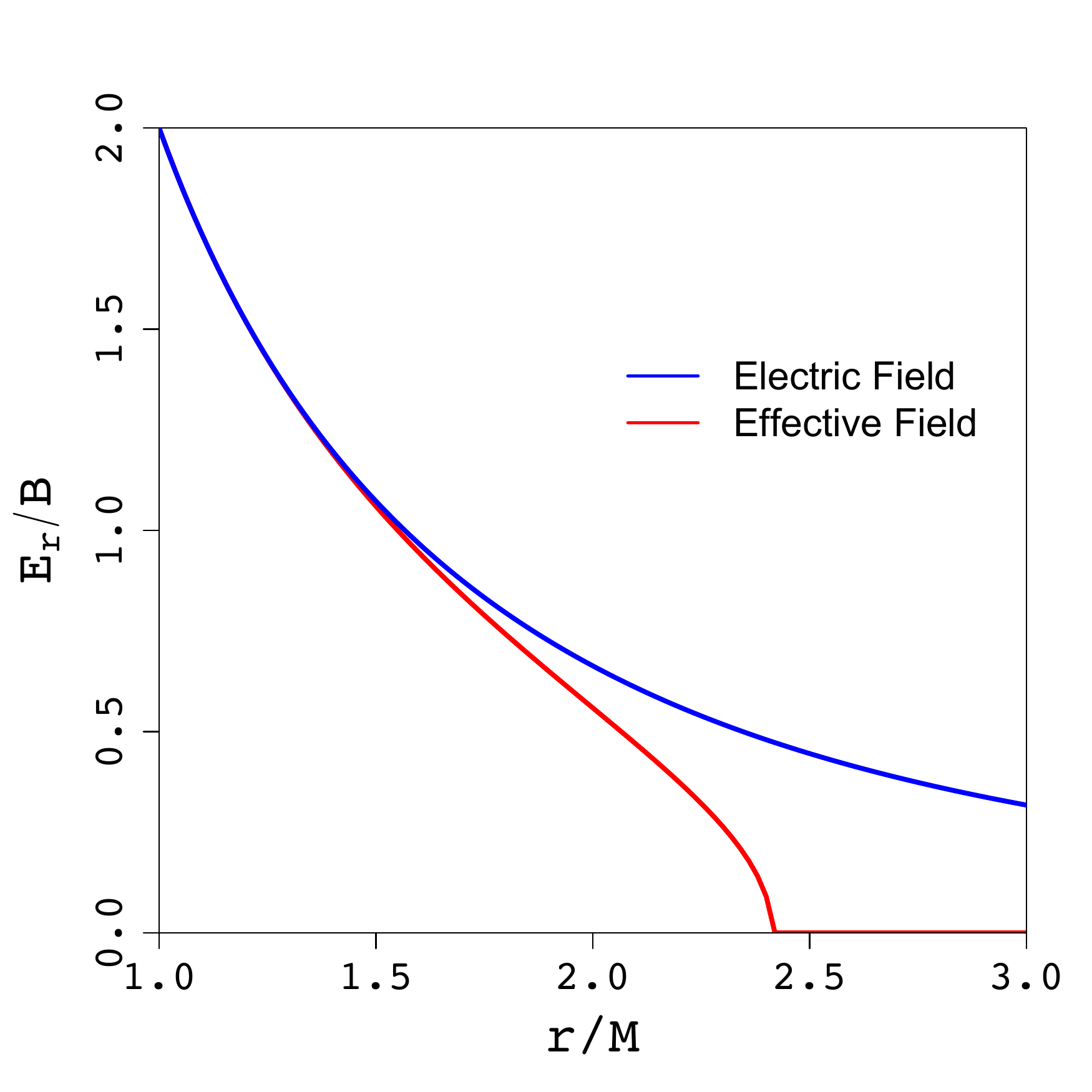}}
\end{center}
\caption{The electric field strength $E_r$ in the non-rotating (zero angular momentum) frame and the effective field strength 
$E_s$, inferred from the Schwinger particle production formula. The field strengths are evaluated in the 
equatorial plane for an extremal black hole.
}
\label{Field}
\end{figure}

Comparing the exponents in the particle production rates Eq. (\ref{srate}) and Eq. (\ref{bhB}) suggest that 
there is an `effective' Schwinger process electric field, 
which we denote by $E_s$,
\begin{equation}
E_s=\hat\omega B
\end{equation}
Figure \ref{Field} shows a comparison between the actual electric field strength, $E_r$, in the locally 
non-rotating frame and the field strength $E_s$ inferred by actual rates. The two agree at the horizon, 
but as we move away from
the horizon the flat space Schwinger result overestimates the production rate.

Although the Schwinger process in flat space does not give the exact exponent, it should still be accurate enough for
calculating the pre-factor in the particle production rate, especially if we use the effective field strength $E_s$
in the Schwinger result (\ref{srate}). The particle production rate per unit proper volume and time should therefore be
\begin{equation}
\Gamma={1\over 4\pi^3}\left({eB\hat\omega\over \hbar}\right)^2e^{-\pi B_0/\hat\omega B}
\end{equation}
The particle production depends on radius $r$ and angle $\theta$. The electrons and positrons move in
circular orbits at near-light speed and generate  current density $J^\mu$, with
\begin{equation}
dJ^\mu=2e\Gamma u^\mu d\tau,
\end{equation}
where $d\tau$ is the proper time interval in the non-rotating frame. The rate of change of
azimuthal current $I$ in the Boyer-Lindquist frame is obtained by a volume integral of $dJ^\phi$,
\begin{equation}
\frac{dI}{dt}=\int_{r_h}^\infty dr \int_0^\pi d\theta\,4\pi\rho^2\sin\theta\,e\Gamma u^\phi
\end{equation}
For a detailed calculation, we can find the value of the 
velocity $u^\phi$ by expanding about the centre of the barrier as before. Consider the velocity components 
$u^\alpha$, where $x^\alpha=\{t,\phi\}$. When
expressed in terms of the velocity, the potential $V$ in Eq. (\ref{Vpot}) becomes
\begin{equation}
V=m^2g_{\alpha\beta}u^\alpha u^\beta+m^2,\label{potv}
\end{equation}
where the $u^\alpha$ velocity components are regarded as functions of $r$ and $\theta$, 
given in terms of the constant momenta by $u^\alpha=g^{\alpha\beta}(p_\beta-eA_\beta)/m$.
We defined the centre of the barrier $r_c$, $\theta_c$ as the point where these functions vanish.
At the ends of the barrier, we use the normal mode $e^i$ from Eq. (\ref{normalmode}) and define
$u^\alpha{}_{,x}=u^\alpha{}_{,i}e^i$,
\begin{equation}
u^\alpha=u^\alpha{}_{,i}\delta q^i=xu^\alpha{}_{,x}\label{azvel}
\end{equation}
The other components $u^i$ vanish at the ends of the barrier by Eq. (\ref{Econs}).
Substituting back into the potential (\ref{potv}) gives
\begin{equation}
u^\alpha=\frac{u^\alpha{}_{,x}}{|u_{,x}|}\label{aznorm}
\end{equation}
where $|u_{,x}|^2=-g_{\alpha\beta}u^\alpha{}_{,x} u^\beta{}_{,x}=\omega^2$.

Finally, we can obtain a rough estimate by expanding $\hat\omega$ about the horizon, 
where $\hat\omega=2$. This gives
\begin{equation}
I\approx \left(\frac{2eBr_h}{\pi\hbar}\right)^2e^{-\pi B_0/2 B}\Delta t
\approx I_0 B_{13}^2e^{-\pi B_0/2 B}\Delta t
\end{equation}
where $B_{13}$ is the magnetic field strength in units of $10^{13}{\rm G}$, 
$I_0=3.4\times 10^{51}{\rm A}$ and $\Delta t$ is the time over which the particles remain in circular orbits. 
The system can only maintain equilibrium if the magnetic field generated by this
current is smaller than the external field $B$.
The induced field near the horizon $B_I\approx \mu_0I/2r_h$.
Requiring $B_I<B$ gives
\begin{equation}
B\lesssim 1.0\times 10^{12}{\rm G}
\end{equation}
Note that this is rather less than the field $B_0=4.4\times 10^{13}{\rm G}$. Nevertheless, the energy of the particles from 
Eq. (\ref{EandL}) is still very high, $E\approx3.00\times 10^{20}\,B_{13} r_{\rm km}\,{\rm eV}$ for 
particles produced close to the horizon $r_{\rm km}$ measured in kilometers.

\begin{figure}[htb]
\begin{center}
\scalebox{0.4}{\includegraphics{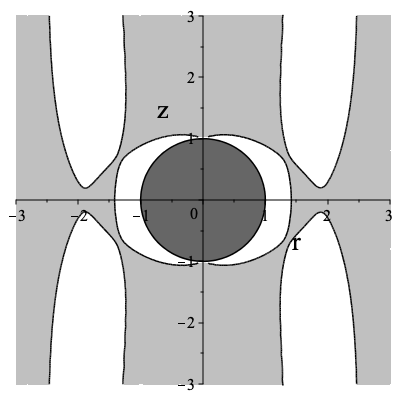}}
\scalebox{0.4}{\includegraphics{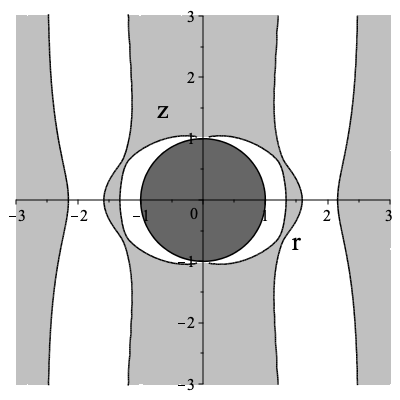}}
\end{center}
\caption{A region of the $r$ and $\theta$ plane showing positive potential (grey) and negative potential (white).
Particles are confined to a white region and produced at the edge. On the left $r_c=1.5M$ and the outgoing particle
cannot cross the equatorial plane. On the right $r_c=1.4M$ and the particle can cross the equatorial plane.
}
\label{Pots}
\end{figure}

We can obtain information about the trajectories 
by looking at the potential diagrams in \ref{Pots}. Initially, due to the closeness to the point $r_c$, $\theta_c$ 
where the potential gradients vanish, the forces moving the particles away from the circular orbits are very small. 
From the potentials, we see that 
the particles produced on the inner edge of the instanton always end up  inside the hole. Particles produced 
on the outside edge can eventually move off to infinity. Depending on the
initial radius, the particle may cross the equatorial plane, but the ones that avoid the 
equatorial plane drift away in the direction along the axis of rotation.

The electrons are produced in high energy circular orbits and should be 
significant sources of synchrotron radiation. In flat space,
the synchrotron emission rate $dE/dt\propto \gamma^2$ where $\gamma$ is the Lorentz factor.
In curved space, the Lorentz factor $\gamma\sim u^t$ and we saw earlier that the tunnelling
process requires special values of the energy and angular momenta, and these imply $u^t\sim 1$.
Consequently, synchrotron emission in the circular orbits is very highly suppressed, and has
a negligible affect on the motion.
However, this only covers the initial circular orbits, and
gradually as the particles drift away from the potential barrier, they will accelerate to higher speeds and 
the emission will increase.

\section{Conclusion}

We have seen some of the tricks employed when using the instanton approach to particle
creation in moderately strong fields and curved spacetimes. It can be viewed as a method
for obtaining quick results from any situation where WKB analysis would be appropriate.
The results are non-perturbative, but the range of usefulness is restricted and prefactors
to the exponential rates are often difficult to obtain.

An example is the rate of vacuum breakdown due to the Schwinger effect 
near a black hole in a magnetic field. The system is very limited because it ignores collisions between 
particles in the surrounding medium, in particular the production from high energy photons
via $\gamma\to e^+e^-$. Nevertheless, it is clear that fields in excess of 
$4.4\times 10^{11}\,G$ will copiously produce electrons of energy above 
$10^{18}{\rm eV}$. The particles produced by the Schwinger mechanism can form a current loop 
around the black hole before drifting off along the rotation axis, whilst
emitting significant synchrotron radiation as they speed up. One application of
the results may be to add particle production terms to relativistic MHD simulations,
using the Schwinger particle production rate in a zero angular momentum frame.

The examples illustrate some interesting features of  the instanton approach to particle creation. 
In particular, the importance of Hamiltonian methods and the distinction between branch cuts
which signal vacuum breakdown and singularities that signal horizon radiation.

\ack This work was supported by the UK Science and Technology 
Facilities Council (STFC) [grant ST/T000708/1].

\section*{References}

\bibliographystyle{iopart-num}
 \bibliography{paper}

\end{document}